\documentclass[12pt]{article}
\usepackage{openwork}
\pdfoutput=1
\usepackage{cite}
\usepackage{authblk}
\usepackage{fancyhdr}
\usepackage{makecell}
\usepackage{adjustbox}
\usepackage{caption}
\usepackage{calc}
\usepackage{chngcntr}
\usepackage{algorithmic}
\usepackage{xcolor}
\usepackage{etoolbox}   
\usepackage[numbers, square]{natbib}
\usepackage[hidelinks]{hyperref}

\counterwithin{figure}{section} % 图编号跟随节重置

\def\BibTeX{{\rm B\kern-.05em{\sc i\kern-.025em b}\kern-.08em T\kern-.1667em\lower.7ex\hbox{E}\kern-.125emX}}

\begin{document} 

\title{Securing High-Concurrency Ticket Sales: A Framework Based on Microservice}  
\author[1]{Zhiyong Zhang\textsuperscript{*}}
\author[1]{Xiaoyan Zhang\textsuperscript{*}}
\author[1]{Xiaoqi Li}
\affil[1]{Hainan University, Haikou, China}
\date{ }

\begingroup
\let\newpage\relax
\thispagestyle{empty}  % 隐藏标题页页码
\setcounter{page}{0}   % 将页码计数器设为0
\maketitle
\endgroup
\setcounter{page}{1}   % 正文从页码1开始
\thispagestyle{fancy}

\begin{abstract}

The railway ticketing system is one of the most important public service infrastructure. In peak periods such as holidays, it is often faced with the challenge of high concurrency scenarios because of a large number of users accessing at the same time. The traditional aggregation architecture can not meet the peak user requirements because of its insufficient fault tolerance and low ability. Therefore, the system needs to use microservice architecture for development, and add multiple security methods to ensure that the system can have good stability and data consistency under high concurrency scenarios, and can respond quickly to user requests. This paper introduces the use of B/S architecture and Spring Cloud to design and develop a railway ticket purchase system that can maintain stability and reliability under high concurrency scenarios, and formulate multiple security design methods for the system. This system integrates a range of functions, such as real-time train inquiries, dynamic seat updates, online seat selection, and ticket purchasing, effectively addressing common problems associated with offline ticket purchasing, such as long queues and delayed information. It enables a complete online process from inquiry and booking to payment and refunds. Furthermore, the "add passenger" function allows users to purchase tickets for others, extending the convenience of online ticketing to people with limited internet access. The system design prioritizes security and stability, while also focusing on high performance, and achieves these goals through a carefully designed architecture and the integration of multiple middleware components. After the completion of the system development, the core interface of the system is tested, and then the results are analyzed. The test data proves that the system has good ability and stability under high concurrency.


\end{abstract} 

\setlength{\parskip}{12pt}
\setlength{\parindent}{2em}
\raggedbottom  % 防止垂直间距不均

\pagestyle{fancy}
\vspace{12pt}
\section{1 INTRODUCTION}

As a critical public service system, the railway ticketing system must handle a massive volume of daily user requests, especially during peak periods when concurrent access can reach hundreds of millions of times. The system processes millions of transactions every day, including not only ticket purchases but also changes and cancellations.

With the rapid popularization of online ticket purchase, more and more users are choosing to book tickets online, which puts greater pressure on the backend system. The traditional monolithic system architecture packages all functions together, and when faced with a large number of users grabbing tickets at the same time, the page often freezes, the operation is slow, and even the same ticket is sold repeatedly. This paper adopts a microservice architecture to build the railway ticketing system\cite{3}, that is, to split the entire system into multiple independent modules such as ticket service, payment service, and order service. Multiple security mechanisms are added to ensure that the system runs smoothly and the data is not corrupted. Each module can be upgraded and maintained separately, so the overall scalability of the system is better. For example, when encountering peak ticket grabbing during holidays, the number of servers for the ticket service can be increased separately. Through this distributed deployment and traffic sharing method, the instantaneous high access volume can be effectively coped with\cite{4}.

The security of a ticketing system is built upon two core requirements: first, the ability to handle massive user traffic during normal operation, and second, maintaining data consistency across all nodes in the event of unexpected failures\cite{3}. A single error can trigger a chain reaction of problems. For example, payment failures can lead to duplicate charges or missed payments, while delays in ticket inventory synchronization can result in overselling. These risks highlight the importance of building distributed ticketing systems based on a microservices architecture and rigorously reviewing their security. This is especially crucial during peak periods such as the annual Spring Festival travel rush, when the system must reliably support billions of ticket requests daily with near-perfect stability.

Researchers and practitioners worldwide have made significant progress in the design and security of railway ticketing systems, particularly through the adoption of microservices architecture and related protective measures. To cope with peak demand, modern systems deploy distributed microservices, combined with caching and message queues, to reduce server load, accelerate response times, and minimize system downtime, ultimately providing users with a smoother experience\cite{5}.

The application of microservice architecture has been widely validated in major railway systems worldwide. For example, Amtrak, facing immense access pressure, rebuilt its online ticketing system using a microservice architecture\cite{7}, deploying instances through cloud services and integrating technologies such as API gateways\cite{8} and Spring Cloud\cite{9}. This transformation significantly improved system stability during peak hours and enhanced the user experience. Similarly, European railway systems\cite{10}, which must handle complex cross-border, high-concurrency scenarios, have adopted service meshes for service governance and utilized Kubernetes for containerized management\cite{11}.

This study designed and developed a stable and secure railway ticketing system based on a microservice architecture. By building an efficient online ticketing platform, it meets people's needs for convenient travel, allowing users to complete ticket inquiries, reservations, payments, and cancellations without leaving home. The system includes functions such as train schedule inquiry, real-time updates of available tickets, and online seat selection, solving problems such as long queues and information delays associated with offline ticket purchases. Furthermore, the system adds features such as allowing users to purchase tickets for multiple passengers, enabling those without internet access to enjoy convenient ticketing services.

This paper conducts an in-depth study of the system's backend framework and security design to ensure system stability, and elaborates on the implementation process of key functional modules. The research includes theoretical analysis, technical design, implementation of core functions (including security measures), and comprehensive performance testing. In the system design phase, a system functional requirements structure diagram was designed based on application needs, technology selection was determined, and a system architecture diagram was drawn. Simultaneously, the system's security design scheme was analyzed, a database E-R diagram was designed, and the data table design was improved based on business relationships. In the system implementation phase, the implementation of the main functional modules is described, the system's security design scheme is analyzed in detail, and performance testing was conducted on the system's core interfaces. The main research content and innovations of this paper are as follows.

(1) Design a microservice-based and high-availability traffic control scheme. The system, which comprises five modules (membership, ticketing, order, payment, and gateway), uses Nacos for their management. Furthermore, Sentinel is introduced to achieve precise circuit breaking, degradation, and interception of abnormal traffic, thereby improving system availability.

(2) Implement a packaged cache component library, use Bloom Filter plus Redis to solve the cache penetration problem, intercept 99\% of invalid requests, and reduce the core interface response time from 200ms to 30ms.

(3) The algorithm for generating distributed globally unique IDs is optimized by proposing to replace UUID with the Snowflake algorithm, which is customized according to the specific scenario within the system. This achieves ordered and monotonically increasing ID generation, is more index-friendly for databases, and improves write performance. Furthermore, tests show that the Snowflake algorithm has better generation performance, higher uniqueness, and smaller storage space.

(4) A method is proposed to monitor MySQL's Binlog logs and then use Canal to capture data change events in real time. Combined with message queues to trigger asynchronous cache updates, a highly efficient and reliable database and cache consistency guarantee system was constructed. Canal parses the Binlog in Row mode, accurately identifies key operations such as order status changes and inventory deductions, encapsulates them into JSON format messages, and sends them to the specified Topic in RocketMQ for asynchronous processing by consumers.

(5) By splitting the traditional monolithic architecture, making reasonable technology selections and adopting a variety of system stability optimization schemes, this system can maintain good stability and performance under high concurrency. The core interface throughput of this system is 817 requests per second and the average response time is 31 milliseconds.

\section{2 BACKGROUND}

\subsection{2.1 Microservices}

\subsubsection{(1) Spring}
\vspace{12pt}

The Spring framework\cite{12} is the basic architecture for Java enterprise application development. Its core principle is based on the two major features of inversion of control and dependency injection. It manages components through the Core Container, which means that after correct configuration, the container will obtain the right to create instance objects and then automatically inject them when needed. At the same time, it separates cross-cutting concerns\cite{13} by means of aspect-oriented programming. For example, logging and transactions are implemented in a non-intrusive way through aspect-oriented programming. Its modular design covers sub-projects such as Spring MVC\cite{14}, Spring Data and Spring Security. Although Spring is a lightweight container, the complexity of the project is high because explicit configuration is required in XML or using Java annotations before developing business code. Therefore, the requirements for developers are very high. Figure 2.1 is the architecture diagram of the Spring framework.

    \noindent
        \begin{center}
        \includegraphics[width=0.6\linewidth]{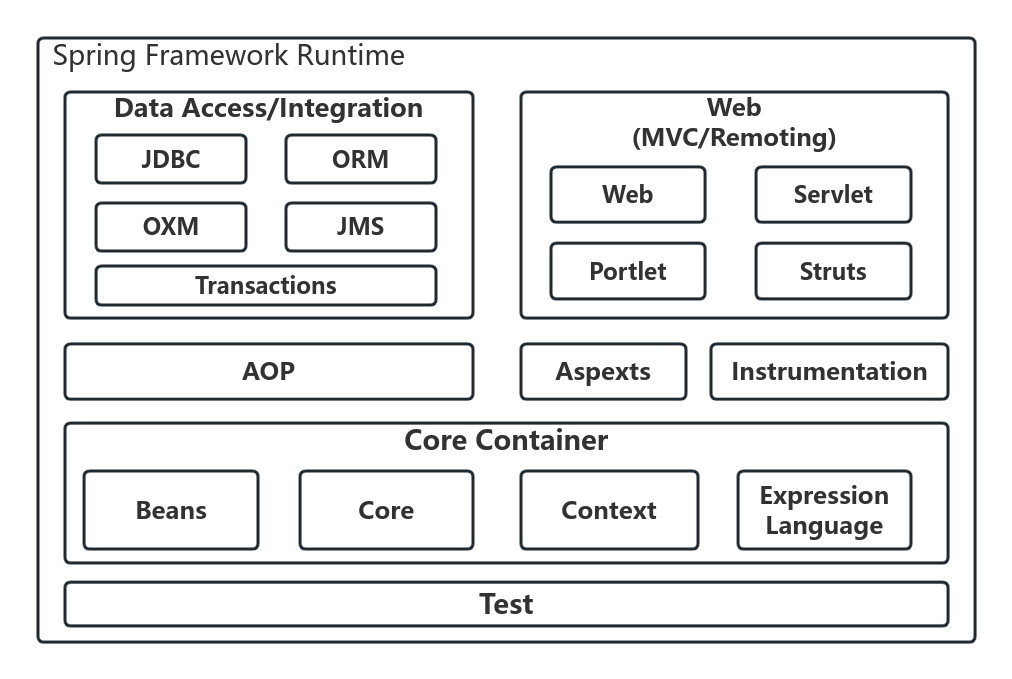}
        \captionof{figure}{Spring Framework Architecture Diagram}
        \label{fig:1}
        \end{center}

\subsubsection{(2)Spring Boot}
\vspace{12pt}

The Spring framework\cite{15} requires a lot of configuration, which can easily lead to various exceptions and low development efficiency. Spring Boot adopts the concept of "convention over configuration" and optimizes the conventional development model. Its core technologies include an auto-wiring mechanism based on conditional configuration, starter dependencies for standardized dependency management, and embedded Servlet containers (such as Tomcat/Jetty).

Auto-configuration is the core function of the Spring Boot framework. After importing the corresponding dependency resource package using a dependency management tool such as Maven\cite{16}, it can be managed through auto-configuration. Figure 2.2 shows the execution flowchart of the Spring Application.

    \noindent
        \begin{center}
        \includegraphics[width=0.8\linewidth]{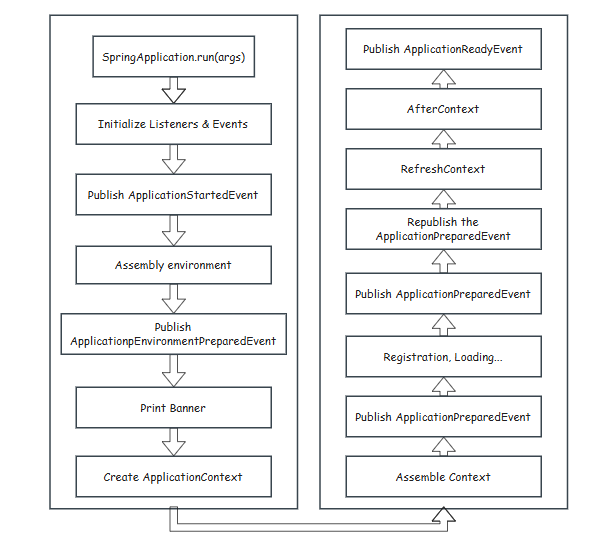}
        \captionof{figure}{Spring Application Execution Flowchart}
        \label{fig:1}
        \end{center}
         

This framework significantly lowers the barrier to entry for Spring applications by simplifying environment configuration and reducing boilerplate code. It also adapts to the microservice architecture requirements of the cloud-native era, becoming the mainstream industry standard for rapidly building production-grade applications and demonstrating significant engineering value in DevOps and continuous integration scenarios.

\subsubsection{(3) Microservice Decomposition Based on Spring Boot} 
\vspace{12pt}

Traditional monolithic architecture integrates various business functions into a single project and deploys them in a unified package. Complex application architecture can lead to poor system availability and impact system performance. Figure 2.3 illustrates a traditional monolithic architecture.

        \noindent
        \begin{center}
        \includegraphics[width=0.85\linewidth]{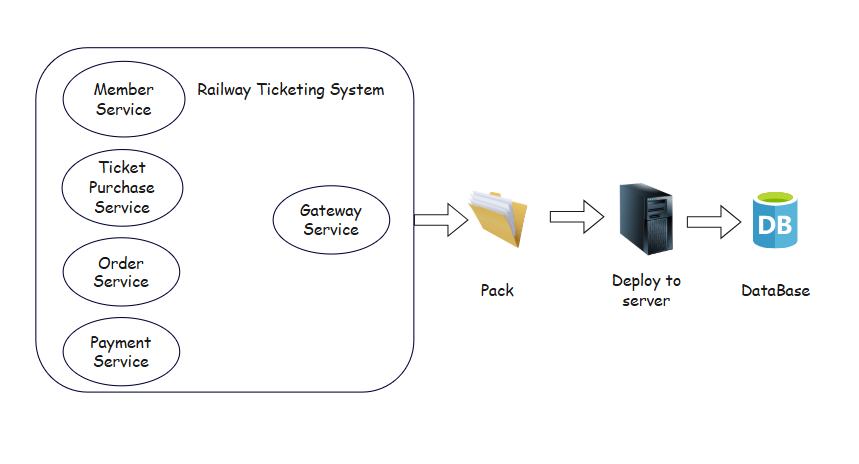}
        \captionof{figure}{Schematic Diagram of Monolithic Architecture}
        \label{fig:1}
        \end{center}

Microservice architecture\cite{17} is a set of best practice solutions guided by the service-oriented approach. It is not a framework itself, but a software architecture style. It is based on many small projects with single functions and responsibilities. By combining them, a complex large application can be realized. Service-oriented architecture is to break down the functional modules in the monolithic architecture into multiple independent projects. Figure 2.4 is a schematic diagram of microservice architecture.

        \noindent    
        \begin{center}
        \includegraphics[width=0.85\linewidth]{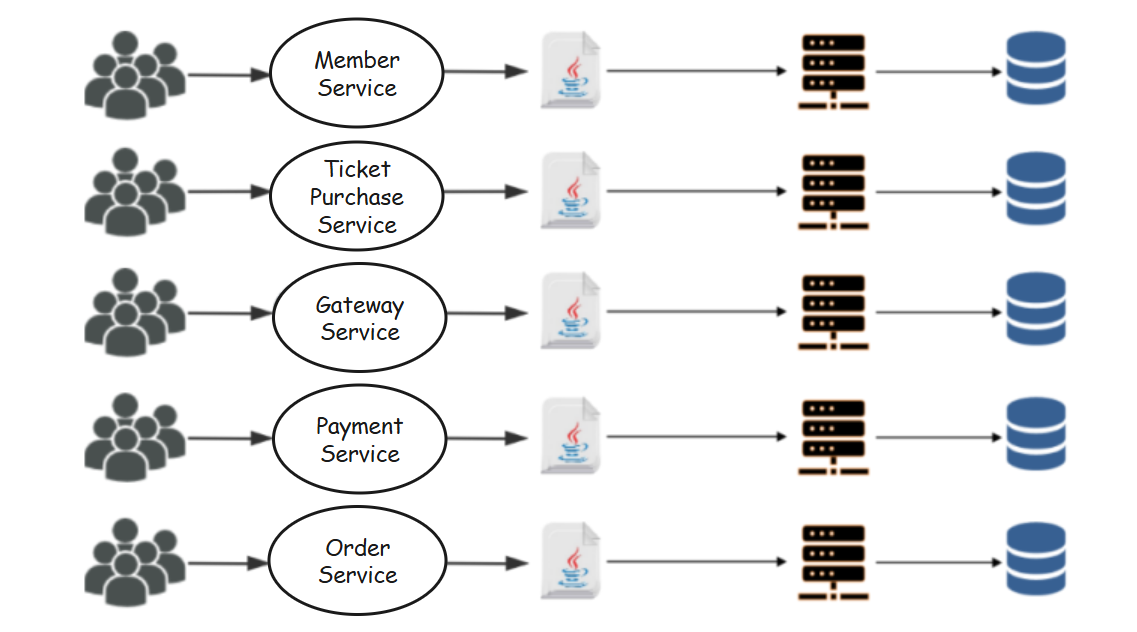}
        \captionof{figure}{Schematic Diagram of Microservice Architecture}
        \label{fig:1}
        \end{center}

Although microservice architecture can improve system availability\cite{18}, it also brings many problems such as how services can make remote calls, how configurations can be uniformly managed, and how applications can monitor traffic. To solve these problems, Spring Cloud Alibaba technology, which is based on the Spring Cloud software architecture style and includes multiple core components, can be used. For example, Nacos can register services and configurations on the server and then manage them, Sentinel can restrict external traffic from entering the system, and RocketMQ can process distributed messages, which helps developers quickly build Java applications with strong availability and good scalability. Spring Cloud Alibaba\cite{19} is compatible with Spring Cloud\cite{20} native components such as OpenFeign and Gateway, which greatly simplifies the workload of developers and has become a popular choice for enterprise-level microservice development. As shown in Figure 2.5, Spring Cloud Alibaba integrates various functional components and implements automatic assembly of these components based on Spring Boot, providing a good out-of-the-box experience\cite{20}.

        \noindent
        \begin{center}
        \includegraphics[width=0.85\linewidth]{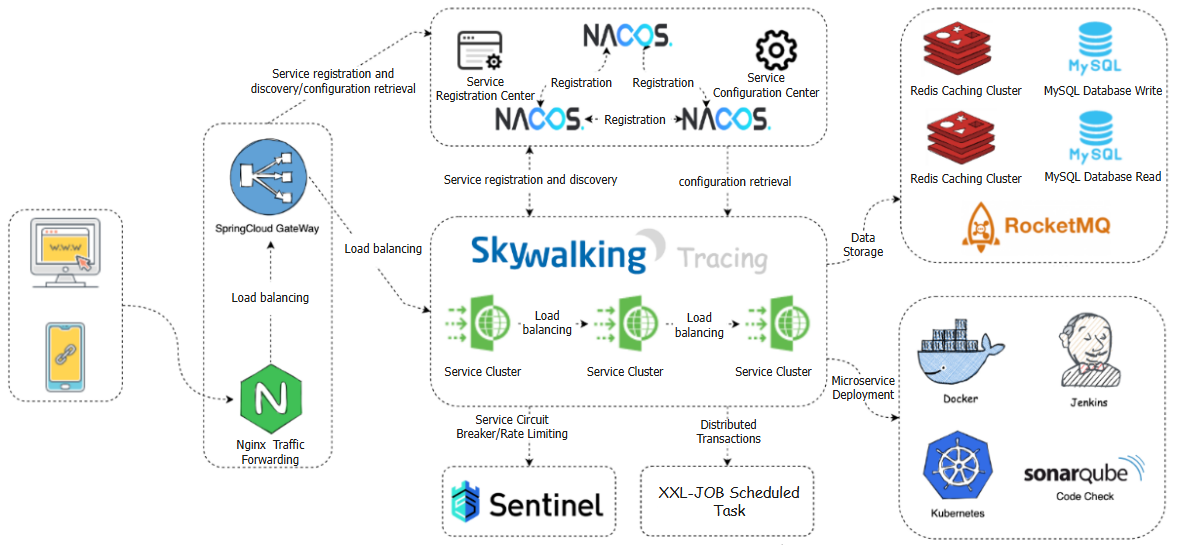}
        \captionof{figure}{Schematic diagram of Spring Cloud Microservice Functional Components}
        \label{fig:1}
        \end{center}

\subsection{2.2 Database}
\vspace{0pt}
\subsubsection{(1) MySQL persistent layer database}
\vspace{12pt}

MySQL\cite{22} is an open-source relational database management system that uses SQL statements to store data in a table structure and supports complex transaction processing and relational queries. Its core features include transaction support, data persistence, and scalability such as support for master-slave replication, database sharding and table partitioning, and read-write separation. These core features are the basis for its ability to cope with high concurrency scenarios\cite{23}.

\subsubsection{(2) Redis cache database}
\vspace{12pt}

Redis is a high-performance in-memory key-value database\cite{23} that responds to read and write operations with extremely low latency. As shown in Figure 2.6, the key is a string type, while the value supports a variety of data types\cite{25}.

    \noindent    
        \begin{center}
        \includegraphics[width=0.85\linewidth]{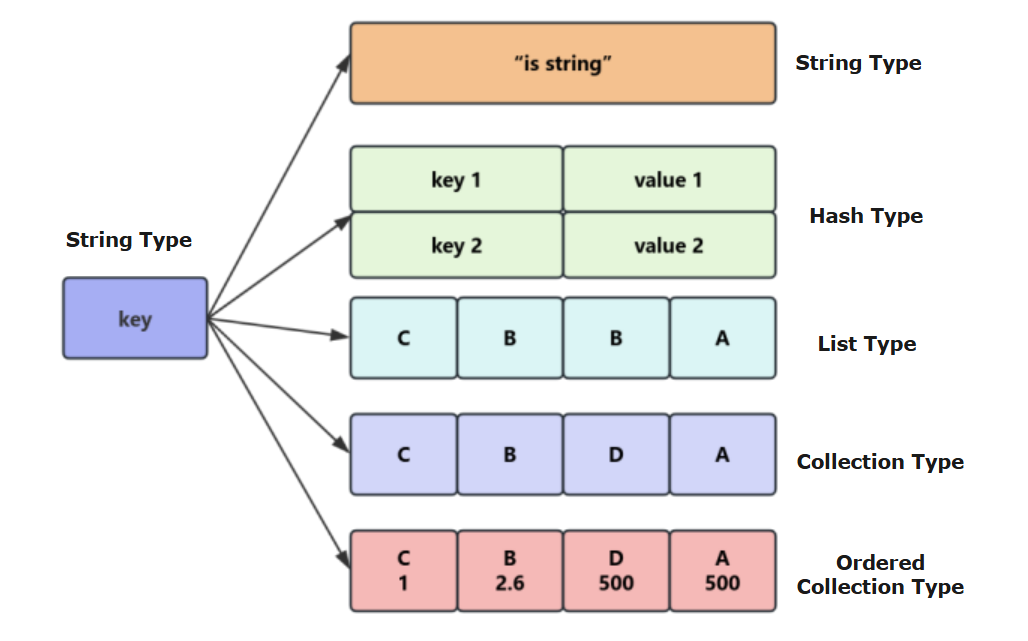}
        \captionof{figure}{Redis Data Types}
        \label{fig:1}
        \end{center}
        
Redis is used as a caching layer to intercept high-frequency requests and reduce the load on the MySQL database. The MySQL database serves as the persistent layer for core data storage (as shown in Figure 2.7). This combination fully leverages MySQL's strong consistency and Redis's high performance to build a reliable and responsive application system\cite{25}.

        \noindent    
        \begin{center}
        \includegraphics[width=0.85\linewidth]{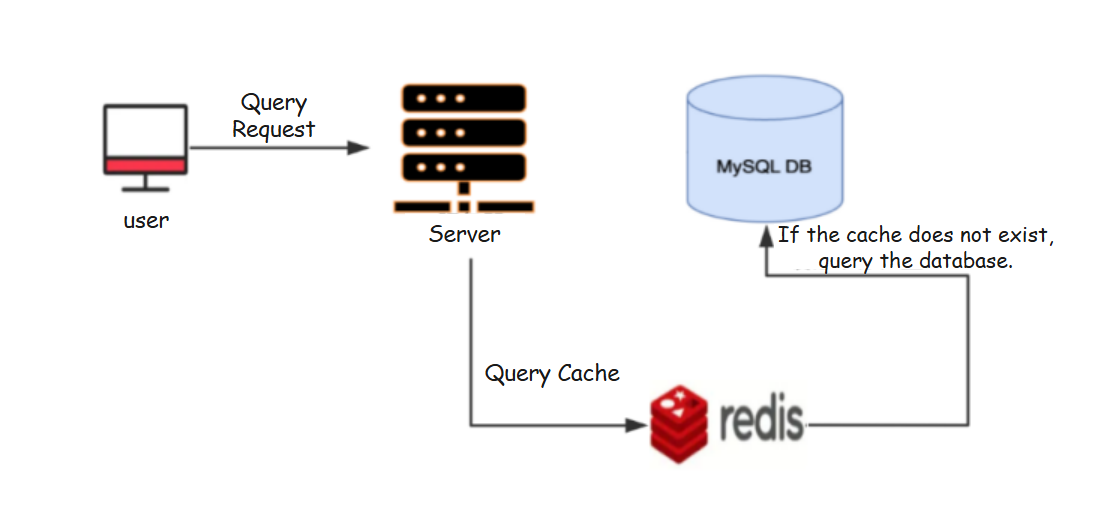}
        \captionof{figure}{Schematic Diagram of MySQL and Redis Collaboration}
        \label{fig:1}
        \end{center}

\subsection{2.3 Canal monitoring}
\vspace{0pt}

Canal is middleware used for real-time monitoring of database log changes. It parses MySQL's Binlog logs and efficiently pushes change events to downstream systems, such as caches, message queues, and big data platforms, thereby achieving real-time synchronization between databases and heterogeneous data sources\cite{27}.

Canal acts as a MySQL replica, reading and parsing the master's Binlog logs to capture data changes. It supports full and incremental synchronization, and is suitable for scenarios such as heterogeneous data replication, cache updates, real-time analysis, and cross-data center synchronization. Canal features low invasiveness, high real-time performance, and high availability, and can be seamlessly integrated with systems such as Kafka, RocketMQ, and Elasticsearch. It is an important tool for building real-time data pipelines\cite{27} and ensuring data consistency. Figure 2.8 illustrates its working principle.

        \noindent
        \begin{center}
        \includegraphics[width=0.85\linewidth]{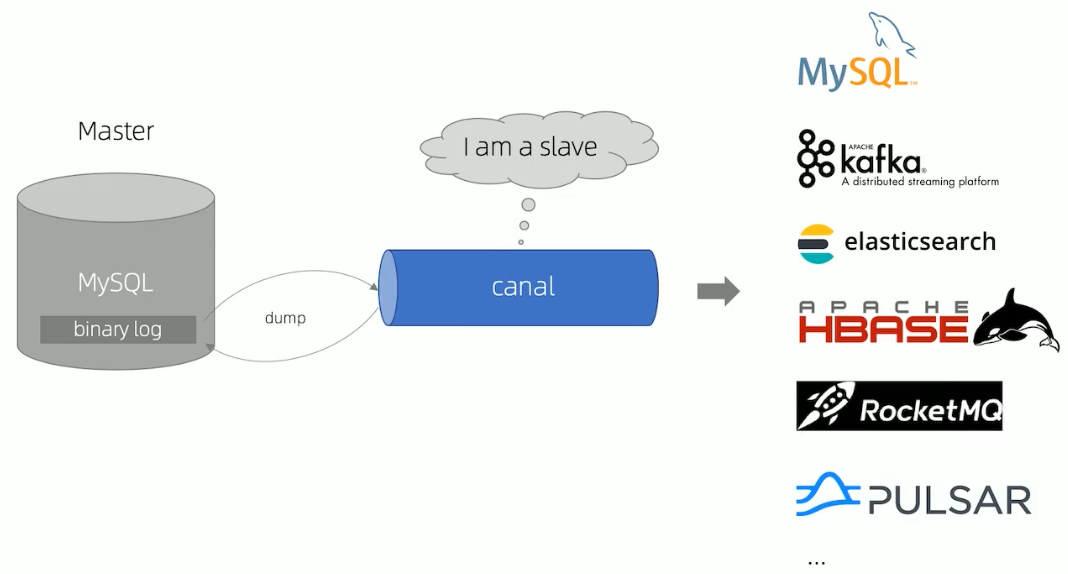}
        \captionof{figure}{Schematic Diagram of Canal Working Principle}
        \label{fig:1}
        \end{center}

\subsection{2.4 Algorithms}

\subsubsection{(1) AES}
\vspace{12pt}

This system uses the AES encryption algorithm and the ShardingSphere framework to implement data encryption. The AES encryption process can be divided into several rounds, each encompassing four main steps: SubBytes, ShiftRows, MixColumns, and AddRoundKey. The encryption process begins with an initial round key addition, followed by N rounds (which determined by the key length) of operations including SubBytes, ShiftRows, MixColumns, and AddRoundKey. However, the final round only requires the first three steps.

M is the initial plaintext, K\_0 is the starting key, and C\_0 is obtained by performing the AddRoundKey operation on the initial plaintext, as shown in Formula 2-1. 

\begin{equation}
    C_0 = M \oplus K_0
    \label{eq:placeholder_label}
\end{equation}  

In each round of encryption, SubBytes byte substitution is performed first, then row shifting is achieved through ShiftRows, followed by column obfuscation transformation through the MixColumns step. Finally, the result of that round is generated by XORing the round key with the AddRoundKey operation. As shown in Formula 2-2, the result after the i-th round of encryption is C\_i, the key used in the i-th round is K\_i.

\begin{equation}  
    C_i = \left[ \left( C_{i-1} \xrightarrow[\text{SubBytes}]{} \textit{ShiftRows} \xrightarrow[\text{MixColumns}]{} \right) \oplus K_i \right]    
    \label{eq:placeholder_label}
\end{equation}

As shown in Formula 2-3, C\_N is the final output ciphertext.

\begin{equation}
    C_{N} = \left( C_{N-1} \xrightarrow{\textit{SubBytes}} \textit{ShiftRows} \right) \oplus K_{N}  
    \label{eq:placeholder_label}  
\end{equation}

SubBytes performs S-Box permutation on each byte as shown in Formula 2-4.

\begin{equation}
    b' = \mathrm{S\mbox{-}Box}(b) \quad (1 \leq b \leq 256)    
    \label{eq:placeholder_label}
\end{equation}

MixColumns matrix operations (finite field GF($2^8$)) are shown in Formula 2-5.

\begin{equation}
    \label{eq:placeholder_label}  
    \begin{bmatrix}
        c_0' \\
        c_1' \\
        c_2' \\
        c_3'
    \end{bmatrix}
    =
    \begin{bmatrix}
        02 & 03 & 01 & 01 \\
        01 & 02 & 03 & 01 \\  
        01 & 01 & 02 & 03 \\
        03 & 01 & 01 & 02
    \end{bmatrix}
    \times
    \begin{bmatrix}
        c_0 \\
        c_1 \\
        c_2 \\
        c_3
    \end{bmatrix}
\end{equation}

The coefficients 02, 03, and 01 are hexadecimal polynomials in GF($2^8$). The multiplication rule here is finite field multiplication, and the addition rule uses the XOR operation.

The $\oplus$ symbol in Formula 2-6 indicates that the operation is performed according to the byte XOR rule, that is, the state matrix and AddRoundKey are XORed by bytes, as shown in Formula 2-6.

\begin{equation}
    \textit{State} = \textit{State} \oplus \textit{RoundKey}  
    \label{eq:placeholder_label}      
\end{equation}

\subsubsection{(2) Customizable Snowflake Algorithm}        
\vspace{-12pt}

There are many scenarios in distributed systems where globally unique IDs are used. To avoid ID conflicts, the Snowflake algorithm can be used\cite{29}. The algorithm generates auto-incrementing long values with customizable lengths, and the ordered auto-incrementing characteristic ensures high performance of B+ Tree index insertion in MySQL\cite{29}. The Snowflake structure is shown in Figure 2.9.

        \noindent
        \begin{center}
        \includegraphics[width=0.85\linewidth]{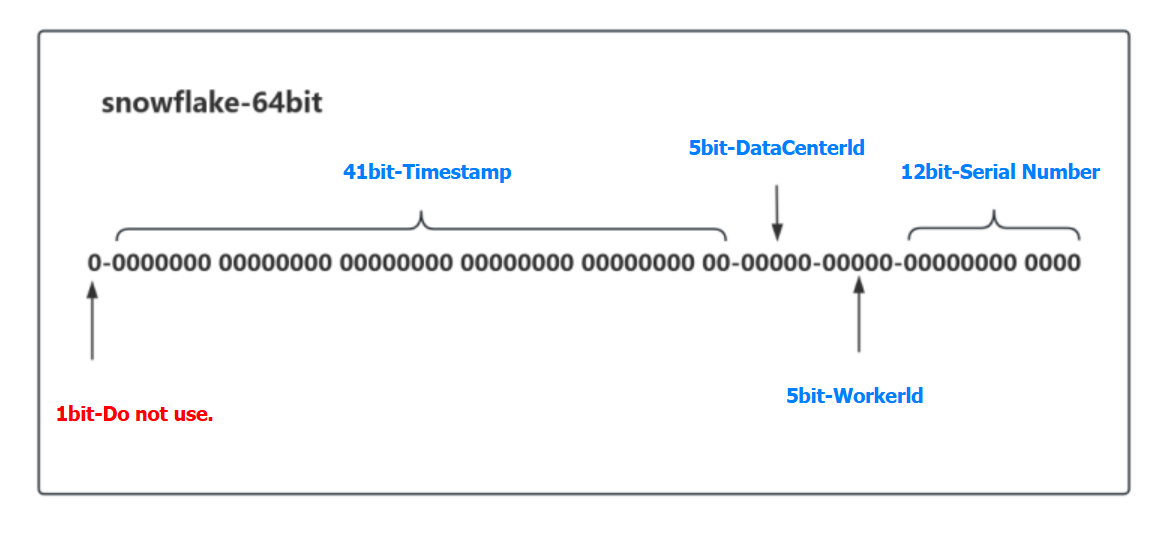}
        \captionof{figure}{Snowflake Algorithm Structure Diagram}
        \label{fig:1}
        \end{center}

The Snowflake algorithm is composed of four parts. The first part is the sign bit, occupying 1 bit. The second part is the timestamp, which uses 41 bits. The third part is the identifier bit, comprising 10 bits in total, with 5 bits allocated to the datacenterId and the other 5 bits to the workerId. These two identifiers are usually combined to represent different deployment nodes. The fourth part is the sequence number, which occupies 12 bits. Because the sequence number is monotonically increasing, it can ensure that the Snowflake algorithm generates globally unique IDs\cite{37}. Then, when the timestamp is fixed, $2^{12}$ unique IDs can be generated per second.

The Snowflake generation formula is based on the OR operation of the displacements of each part, as shown in Formula 2-7.

\begin{equation}
    \begin{aligned}
    \text{long id} &= \big((\text{timestamp} - \text{twepoch}) \ll \text{timestampLeftShift}\big) \\      
              &\quad \mid \big(\text{datacenterId} \ll \text{datacenterIdShift}\big) \\    
              &\quad \mid \big(\text{workerId} \ll \text{workerIdShift}\big) \\        
              &\quad \mid \text{sequence}  
    \end{aligned}
\end{equation}    

The parameter twepoch represents the custom epoch start time. The parameter timestampLeftShift indicates the number of bits for the timestamp left shift. The parameter datacenterIdShift defines the number of bits for the datacenter ID left shift. The parameter workerIdShift specifies the number of bits for the worker machine ID left shift. Finally, the parameter sequence denotes the auto-incrementing sequence within the current millisecond.

\subsubsection{(3) Bloom Filter}
\vspace{12pt}

Bloom Filter utilize bit arrays and multiple hash functions to achieve fast member lookup, making them an efficient and reliable data structure\cite{29}. The working principle of a Bloom Filter is usually to use a binary array to represent a set, with all data in the array initially set to 0. Then, for each element to be added to the container, multiple hash values are calculated using multiple hash functions, and the corresponding positions in the bit array are set to 1. When determining whether an element exists in the container, the element also calculates its corresponding hash value using multiple hash functions, and then checks whether the data stored in the corresponding positions in the bit array are both 1. If both are 1, since hash collisions may occur, the element may only exist in the container, and further querying of the underlying data storage container is needed to confirm whether it really exists\cite{31}. However, if any position is 0, it means that the element definitely does not exist in the container. The principle diagram is shown in Figure 2.10.

        \noindent
        \begin{center}
        \includegraphics[width=0.85\linewidth]{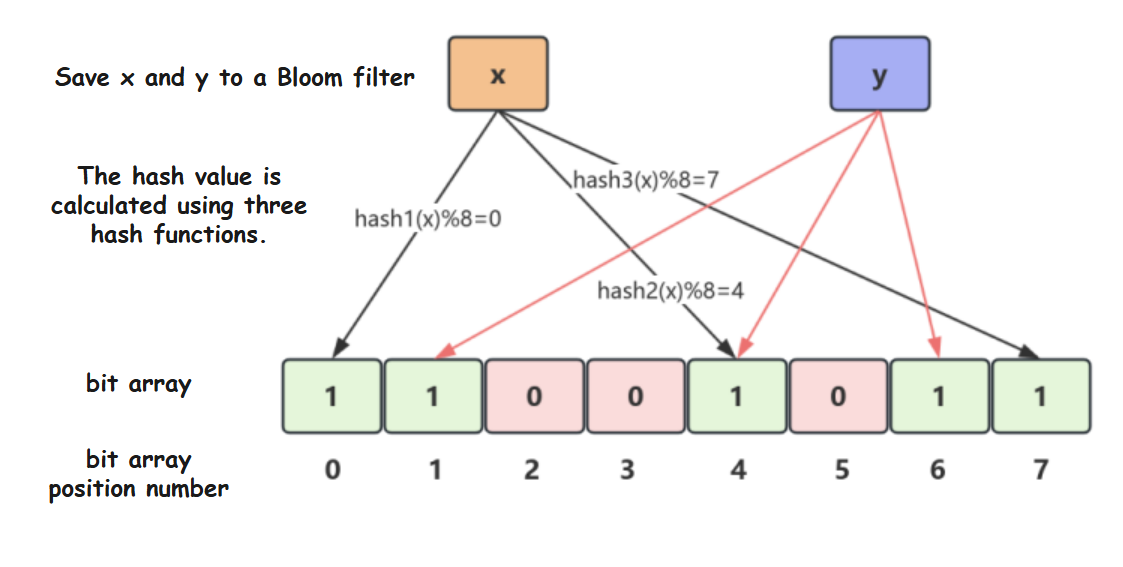}
        \captionof{figure}{Schematic Ciagram of a Bloom Filter}
        \label{fig:1}
        \end{center}

The formula for calculating the false positive rate of a Bloom Filter is shown in Formula 2-8.  

\begin{equation}
    p \approx \left( 1 - e ^ { \frac { k n } { m } } \right) ^ { \kappa }    
    \label{eq:placeholder_label}
\end{equation}

The core parameters are the number of elements n, the size of the binary array m, and the number of hash functions k. Therefore, the required binary array size m for calculating the false positive rate can be derived as shown in Formula 2-9.

\begin{equation}
    m = - \frac{n \ln p}{(\ln 2)^2}      
    \label{eq:placeholder_label}      
\end{equation}

\section{3 SYSTEM DESIGN}

\subsection{3.1 Module Design}
\vspace{0pt}

This system comprises five modules, including the membership, order, ticketing, payment, and gateway modules. Figure 3.1 illustrates the functional requirements structure of this system.

         \noindent
        \begin{center}
        \includegraphics[width=0.7\linewidth]{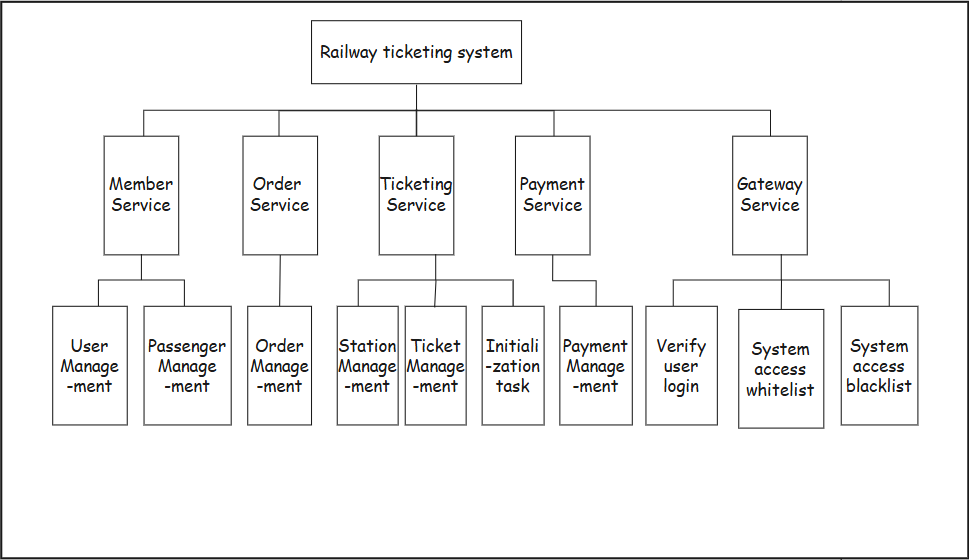}
        \captionof{figure}{Functional Requirements Structure Diagram of Railway Ticketing System}
        \label{fig:1}
        \end{center}

The diagram illustrates the functional composition of each service. The membership service includes user and passenger management. The ticketing service comprises station management, ticket management, and initialization tasks. The order service is dedicated to order management. The payment service handles payment management. Finally, the gateway service provides user login verification, system access whitelisting, and blacklisting. The detailed functions of each module are as follows.

\subsubsection{(1) Membership Module}
\vspace{12pt}

This module is responsible for user identity and passenger management. Its user management functions cover the entire account lifecycle, including checking username availability, new user registration, login (via mobile number, email, or username), login status verification, user information retrieval, and user account deletion. For passenger management, the module provides functions for adding, modifying, and deleting passengers, and supports querying passenger lists based on associated usernames or a set of passenger IDs.

\subsubsection{(2) Ticketing Module}
\vspace{12pt}

This module comprises three core areas, station management, ticketing management, and system initialization. The station management component provides functions for querying station information, including city-based station lists and station details for specific trains. In terms of ticketing operations, it supports conditional ticket searches, online seat selection, and automatic seat inventory management—deducting seats upon successful payment and rolling back seat assignments upon order cancellation. Furthermore, the system periodically runs a series of scheduled initialization tasks to update regional station data, synchronize train route information, and refresh real-time ticket availability for each station.

\subsubsection{(3) Order Module}
\vspace{12pt}

This module is responsible for managing the complete lifecycle of ticketing orders, while ensuring data security and accessibility. When creating an order, it uses a custom generator to generate a unique order number and takes measures to prevent duplicate submissions. Throughout the order processing process, it handles state transitions, such as delaying order closure, restoring the order status after cancellation, and triggering callbacks after payment is completed. To protect sensitive information, the module applies data anonymization to query operations and encrypts order data before storing it in the database. Passengers can also use this module to query their ticketing orders.

\subsubsection{(4) Payment Module}
\vspace{12pt}

This module integrates with Alipay to facilitate secure and convenient payment processing. Its core functions include initiating payment requests, handling successful callbacks from the payment gateway, and providing payment status inquiry capabilities.

\subsubsection{(5) Gateway Module}
\vspace{12pt}

As the entry point of the system, this module implements critical security control functions. It is responsible for verifying user login status and enforcing access policies through whitelisting (allowing access) and blacklisting (blocking access) mechanisms.

\subsection{3.2 Security Design}
\vspace{0pt}
\subsubsection{(1) System Stability}
\vspace{12pt}

In high-concurrency scenarios, traditional monolithic architectures face immense server access pressure, leading to slow response times and even system crashes\cite{32}. Furthermore, malicious request attacks exacerbate these problems, resulting in serious consequences\cite{38}. Therefore, mitigating server pressure and ensuring system stability and performance are crucial.

\subsubsection{(2) Database Security}
\vspace{12pt}

The ticketing system is an application that frequently faces high concurrency scenarios for writing and querying requests\cite{33}. When the application faces a massive number of write requests, a single database server may not be able to handle the huge write pressure. The data can be split and stored in multiple database servers according to the rules set by the user. In this way, each database only needs to handle the write requests of a portion of the data, which alleviates the pressure on the database server and thus improves the write performance. Furthermore, the data in the data table of each database server is split into multiple tables in the same database according to the rules set by the user. Each table only needs to store a portion of the data, which avoids high-frequency operations on only one table in a database server. This solves the problem of decreased storage and query performance of a single data table. This system uses ShardingSphere middleware, which can efficiently and securely perform database sharding and table design. At the same time, when the application faces high concurrency requests, if all requests go directly to the database server, it will greatly increase the pressure on the database server. A cache database is needed to share the pressure of the underlying persistent database. Therefore, a cache layer is designed to improve the system's response speed and enhance the system's stability. And by using a variety of efficient and feasible solutions, the problems of cache breakdown and avalanche caused by using a cache are avoided. 

\subsubsection{(3) Data Consistency}
\vspace{12pt}

To reduce database pressure, this system has designed a caching layer to store frequently queried data in the cache. However, this leads to inconsistencies between the cache and the database\cite{34}. A reliable and secure solution needs to be designed to ensure data consistency between the cache and the database.    

\subsubsection{(4) Overstocking}
\vspace{12pt}

Furthermore, this system needs to ensure that ticket inventory is not oversold when a large number of users purchase tickets simultaneously in high-concurrency scenarios. A reasonable solution is needed to solve this problem while ensuring system performance. 

\subsection{3.3 Architecture Design}
\vspace{0pt}

The system is designed based on B/S architecture, follows the design principle of MVC, and has been thoroughly classified. By implementing a layered strategy, it ensures that each layer focuses on its core task and follows the open/closed rule, thus facilitating adjustment and expansion. Using JDK17+Spring Boot 3+Spring Cloud, a ticketing system that can provide secure and efficient service even when a large number of users access it concurrently was designed and implemented\cite{35}. Figure 3.2 shows the detailed technical structure of the system. 

    \noindent
        \begin{center}
        \includegraphics[width=0.8\linewidth]{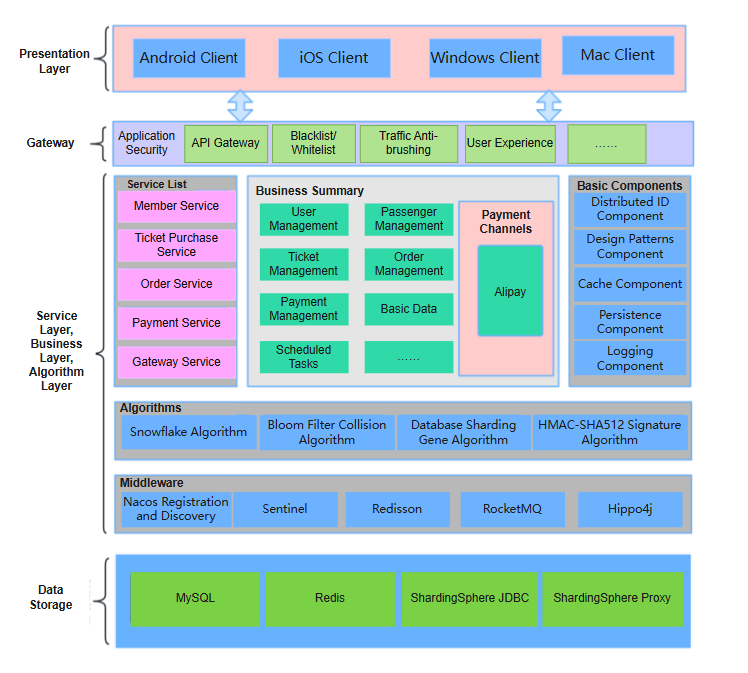}
        \captionof{figure}{Overall Architecture Diagram of Railway Ticketing System}
        \label{fig:1}
        \end{center}

\subsection{3.4 Data design}
\vspace{0pt}

\subsubsection{3.4.1 E-R Diagram}
\vspace{12pt}

The database model structure involved in this system is shown in Figure 3.3.

    \noindent
        \begin{center}
        \includegraphics[width=0.85\linewidth]{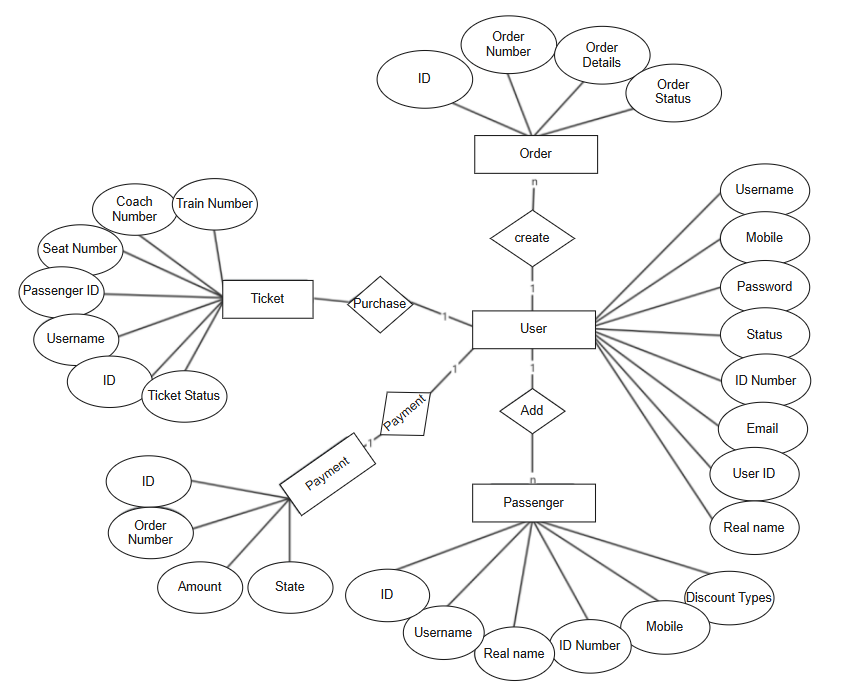}
        \captionof{figure}{E-R Diagram of Railway tTicketing System Database}
        \label{fig:1}
        \end{center}

\subsubsection{3.4.2 Relationship Table}
\vspace{12pt}
\subsubsection{(1) Membership Management}
\vspace{12pt}

\begin{itemize}[noitemsep]
\item t\_user is the member data table, which stores the basic information of account users. 
\item t\_user\_mail is a member email data table that stores the mapping between email addresses and usernames.
\item t\_user\_phone is a member mobile phone number data table that stores the mapping between mobile phone numbers and usernames.
\item t\_user\_reuse is a username reusable table that stores available usernames that have been logged out.
\item t\_user\_deletion is the user ID number deletion table, which stores the data of ID number records that have been deleted.
\item t\_passenger is the passenger data table.
\end{itemize}

The relationships between the various data tables involved in membership management are shown in Figures 3.4 and 3.5. 

    \noindent
        \begin{center}
        \includegraphics[width=0.85\linewidth]{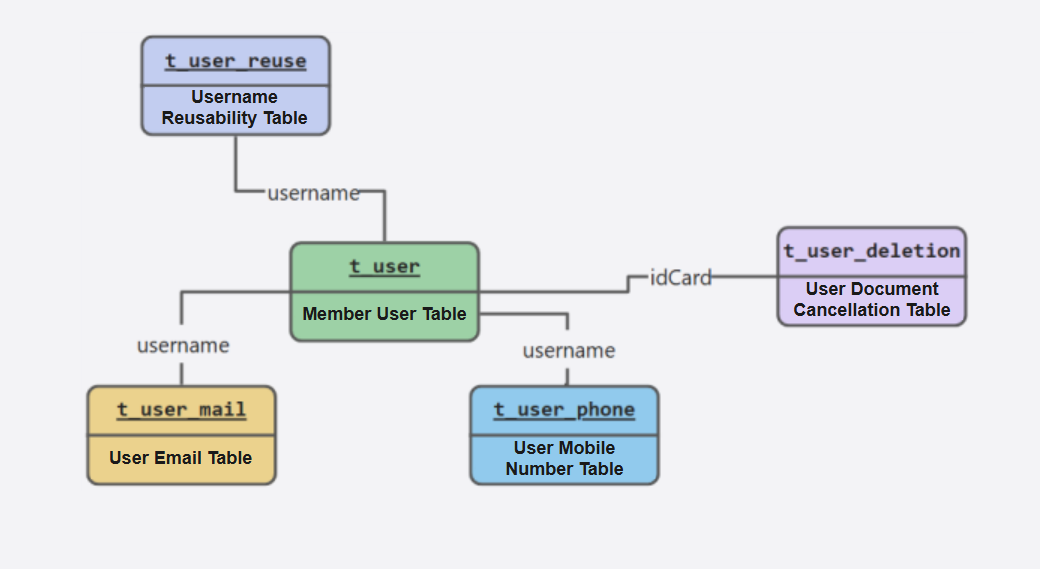}
        \captionof{figure}{E-R Diagram of Railway Ticketing System Database}
        \label{fig:1}
        \end{center}

        \noindent
        \begin{center}
        \includegraphics[width=0.85\linewidth]{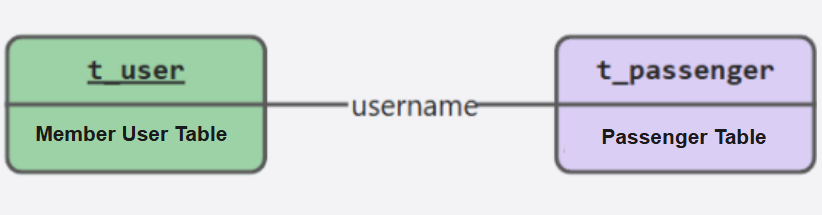}
        \captionof{figure}{E-R Diagram of Railway Ticketing System Database}
        \label{fig:1}
        \end{center}

\subsubsection{(2) Train Management}
\vspace{12pt}
\begin{itemize}[noitemsep, leftmargin=*]
\item t\_train is used to store daily train schedule information.
\item t\_carriage stores the carriage information for all train services.
\item t\_train\_station is a list of train stations.
\item t\_train\_station\_relation is a train station relationship table.
\item t\_train\_station\_price is a price list for train stations.
\end{itemize}

The relationships between the various data tables involved in train management are shown in Figure 3.6.  

    \noindent
        \begin{center}
        \includegraphics[width=0.6\linewidth]{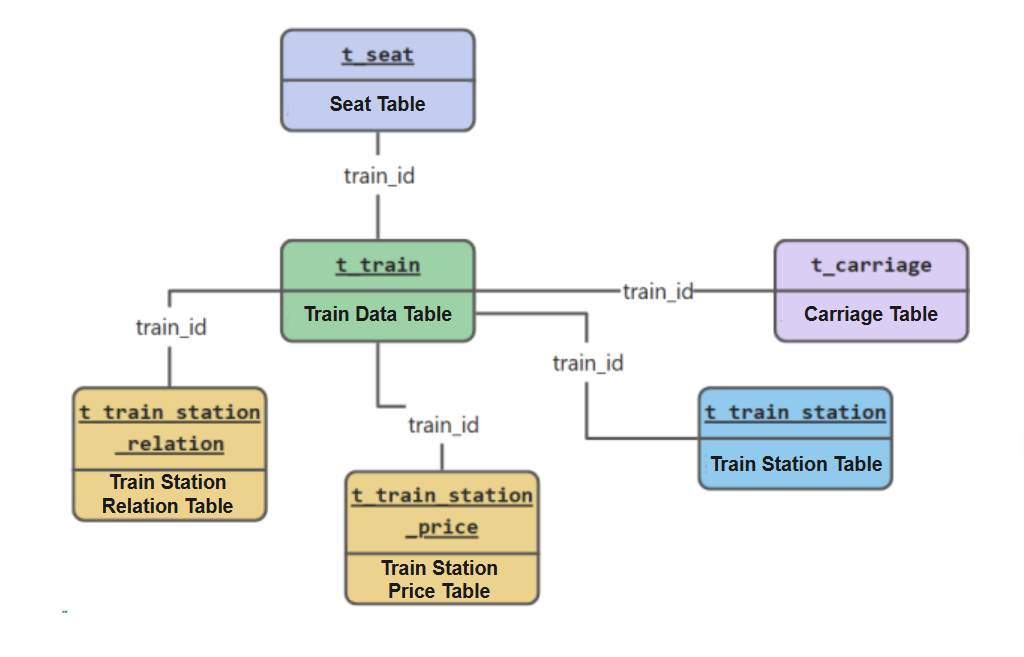}
        \captionof{figure}{E-R Diagram of Railway Ticketing System Database}
        \label{fig:1}
        \end{center}
        
\subsubsection{(3) Order Management}
\vspace{12pt}
\begin{itemize}[noitemsep, leftmargin=*]
\item t\_order represents a main order record corresponding to a user's purchase of a ticket.
\item t\_order\_item represents the number of details corresponding to the number of passengers in the main order.
\item The `t\_order\_item\_passenger` function effectively solves the problem of mismatch between sharding keys and query conditions by associating passenger ID numbers with order details, thereby improving the efficiency and accuracy of data retrieval.
\end{itemize}

The relationships between the various data tables involved in order management are shown in Figure 3.7.  

    \noindent
        \begin{center}
        \includegraphics[width=0.85\linewidth]{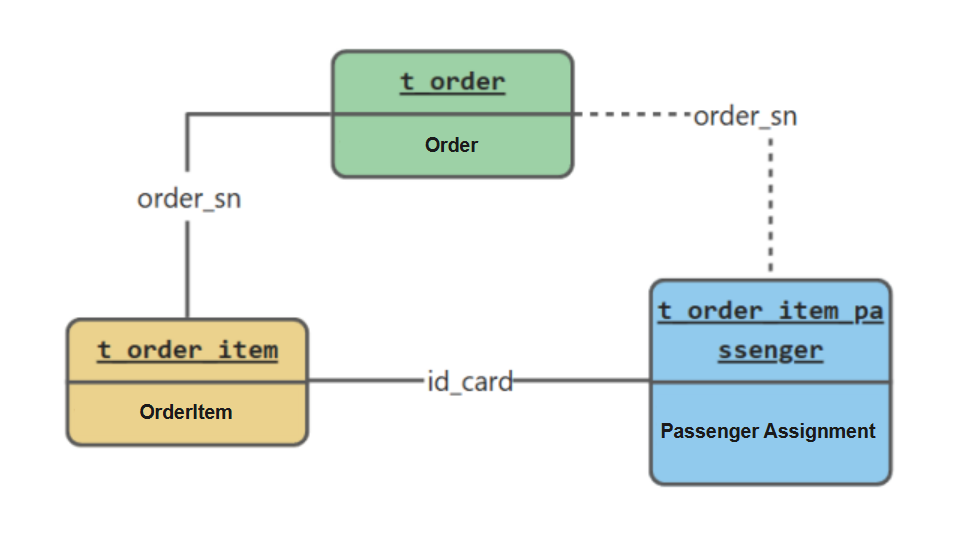}
        \captionof{figure}{E-R Diagram of Railway Ticketing System Database}
        \label{fig:1}
        \end{center}

\subsubsection{(4) Payment Management}
\vspace{12pt}

\begin{itemize}[noitemsep, leftmargin=*]
\item t\_pay is the order payment table, used to store data related to user payments for train tickets.
\end{itemize}

The relationships between the various data tables involved in payment management are shown in Figure 3.8.    

    \noindent
        \begin{center}
        \includegraphics[width=0.5\linewidth]{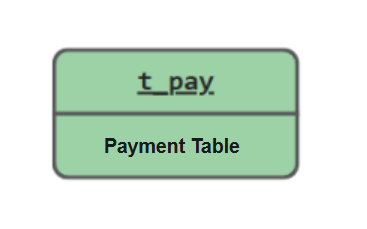}
        \captionof{figure}{E-R Diagram of Railway Ticketing System Database}
        \label{fig:1}
        \end{center}

\section{4 IMPLEMENTATION}
\vspace{0pt}
\subsection{4.1 Security Implementation}
\vspace{0pt}
\subsubsection{4.1.1 System Stability and High Performance}
\vspace{12pt}

The system's business logic is broken down into functional modules. This system is divided into five modules based on business functions, including Membership Service, Ticketing Service, Order Service, Payment Service, and Gateway Service. These microservices are deployed in a distributed manner on different servers using different databases. All registered microservices are managed within the Nacos server using the Nacos component. User requests from the browser are routed and forwarded by the gateway service to the correct microservice on the Nacos server. Inter-service calls are implemented using the OpenFeign component.

Implementing anti-fraud strategies and traffic limiting is crucial for enhancing system security and maintainability. For example, if the order service fails, but the ticketing service continues to send requests, it can waste resources and lead to cascading failures. To prevent this, the system uses Sentinel for traffic control. Sentinel monitors core interface metrics, including queries per second (QPS), response time, and thread count, and dynamically applies rate limiting and circuit breaker strategies to intercept abnormal traffic and protect server resources from exhaustion. Figure 4.1 shows the rate limiting scheme of this system. 

    \noindent
        \begin{center}
        \includegraphics[width=0.85\linewidth]{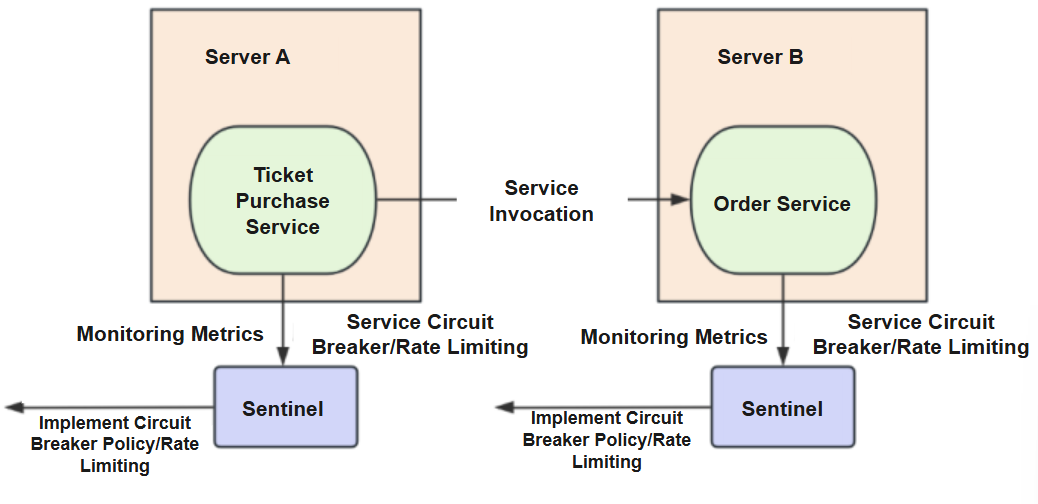}
        \captionof{figure}{Sentinel Rate Limiting Scheme}
        \label{fig:1}
        \end{center}
    
\subsubsection{4.1.2 Database Stability and High Performance}
\vspace{12pt}

When using ShardingSphere for database sharding, choosing the appropriate sharding key is crucial. The system needs to consider several factors when selecting a sharding key, using a user ID can centralize user-related data, but frequent access by multiple users might lead to hotspot issues. On the other hand, using continuously increasing values like order timestamps can lead to uneven data distribution over time. Therefore, an ideal sharding key should possess characteristics such as even distribution, high relevance to queries, and immutability. Based on these factors, this system chose the username field as the sharding key for the user table. User ID and order number are defined as sharding keys. If a query request includes these two fields, the last six digits of the user ID are used to determine the specific location in the sharded table. Since querying passenger data inevitably involves user information, username was chosen as the sharding key for the passenger table. The basic query requirements for order sharding are that users can view their own orders and accurately retrieve order numbers.

Meanwhile, this system uses Redis in-memory key-value database as a caching layer to intercept high-frequency requests, and MySQL as the persistent database. By combining the strong consistency of MySQL with the high performance of Redis, a reliable and responsive railway ticketing system is built. Specifically, frequently queried data, such as user account information, train information, and available tickets, is stored in Redis. When a user accesses this data, Redis can directly return a response, improving response speed and reducing database pressure. 

However, using Redis as a caching layer can lead to cache penetration. This problem may occur when the system receives malicious or high-frequency requests targeting data that does not exist in the cache. Because the cache layer lacks the data, it cannot respond to such requests, and each query directly penetrates to the database, causing the caching mechanism to fail. This bypassing of the cache forces the database to frequently respond to invalid requests, causing resource overload and ultimately severely impacting system performance. For example, in the user registration process of this system, if a new user logs in, they first need to register. After entering their username, the system needs to check if the username is already in use. This request usually requires further access to the database server. When the system determines username availability through both cache and database queries, frequent requests for usernames not existing in the database during high-concurrency scenarios (such as a large number of new user registrations) can lead to the following problems, the cache layer misses the cache and directly penetrates to the database for queries. Since there is no corresponding data in the database, null values are not cached, and subsequent identical requests will continue to bypass the cache layer, forcing the database to repeatedly process invalid queries. This ultimately causes cache penetration, leading to database resource overload, response delays, and even system crashes. This is common in scenarios involving malicious attacks or sudden high-concurrency access to non-existent data. 

These problems are often triggered by cache penetration when facing malicious attacks, frequent access to cache-missing data, or sudden high-concurrency queries after cache expiration. This system utilizes the characteristics of a Bloom Filter to evaluate the data in user query requests. Data that does not exist after filtering by the Bloom Filter is directly returned as non-existent by the system. Only requests for data that exists will proceed to the next query process. The specific solution is shown in Figure 4.2. 

     \noindent
        \begin{center}
        \includegraphics[width=0.85\linewidth]{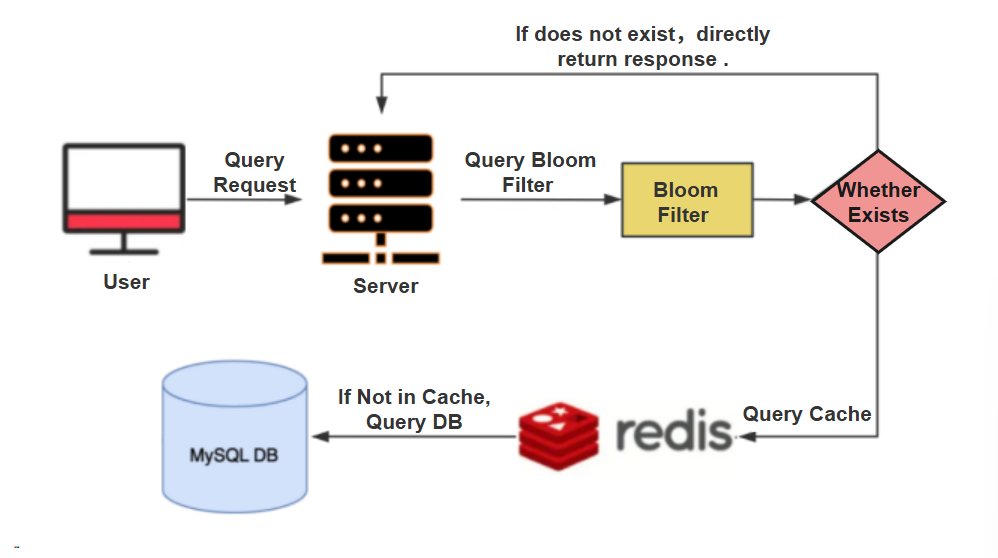}
        \captionof{figure}{Cache Penetration Solution}
        \label{fig:1}
        \end{center}
    
The final solution chosen for this system combines Bloom Filter with caching, which avoids frequent database penetration queries and essentially solves the cache penetration problem. 

\subsubsection{4.1.3 Database and Cache Consistency}
\vspace{12pt}

This system uses Redis as a caching layer. Since Redis caches frequently accessed information such as ticket availability, the system needs to update both the persistent layer and the cache simultaneously when data is updated in actual business operations. Upon receiving a user's ticket purchase request, the persistent database deducts the number of tickets. Because any changes to the data in the database are captured in its binlog file, the Canal middleware is used to capture changes in the MySQL database's binlog file. The captured data changes are then forwarded to a specific topic in a message queue, and the cache is updated asynchronously. Client applications can listen to this message queue topic to maintain consistency between the cache and the database. Figure 4.3 shows the sequence diagram of the database and cache consistency solution.

    \noindent
        \begin{center}
        \includegraphics[width=0.85\linewidth]{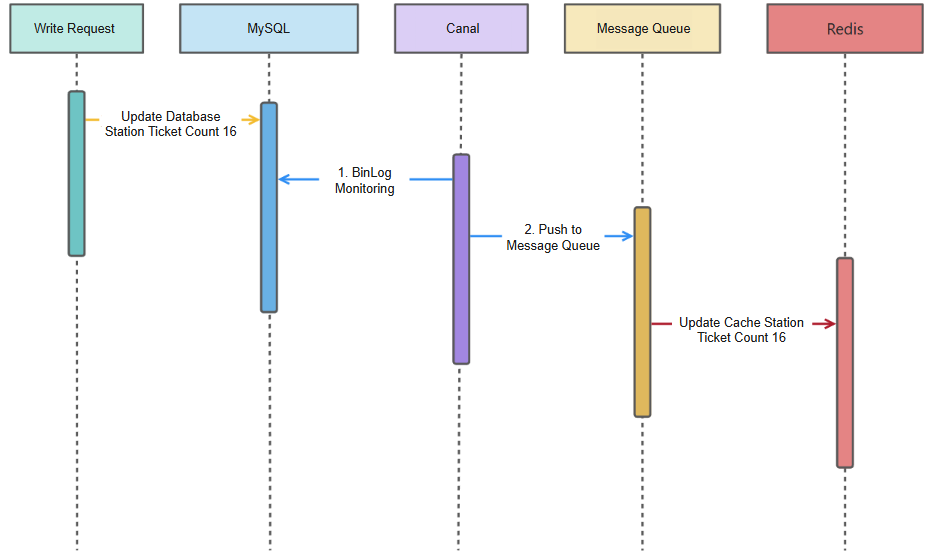}
        \captionof{figure}{Sequence Diagram of Database And Cache Consistency Solution}
        \label{fig:1}
        \end{center}

\subsubsection{4.1.4 Prevent Overselling}
\vspace{12pt}

The system initially addressed data consistency issues between the MySQL database and the Redis cache by monitoring the binlog of the MySQL database using Canal and then asynchronously writing to the cache via a message queue. However, in high-concurrency scenarios such as ticket-buying, each purchase request requires a data consistency mechanism to deduct the remaining tickets from the cache. This leads to a surge in server and database load due to a large influx of requests within a short period. Furthermore, the order in which multiple requests update the remaining tickets in Redis differs from the order in which the remaining tickets are deducted from the database, resulting in overselling. For example, user A checks Redis for 1 remaining ticket, then updates the remaining tickets to 0 in MySQL. Canal monitors this and asynchronously updates the cached remaining tickets to 0 via a message queue. However, before user A's cache update is complete, user B also attempts to buy a ticket. At this point, Redis still shows 1 remaining ticket, allowing user B to proceed with the purchase, resulting in overselling.

This system implements rate limiting and resolves the overselling issue by building an inventory token container from the Redis cache database. Figure 4.4 shows the keys in the split token container. The token container is a Hash structure in Redis, and "Departure Station - Arrival Station - Seat Type" in the figure are all internal keys in the Hash structure. Each key corresponds to a value, which is the remaining amount of "Departure Station - Arrival Station - Seat Type". The tokens stored in it are the train ticket availability data.  

    \noindent
        \begin{center}
        \includegraphics[width=0.85\linewidth]{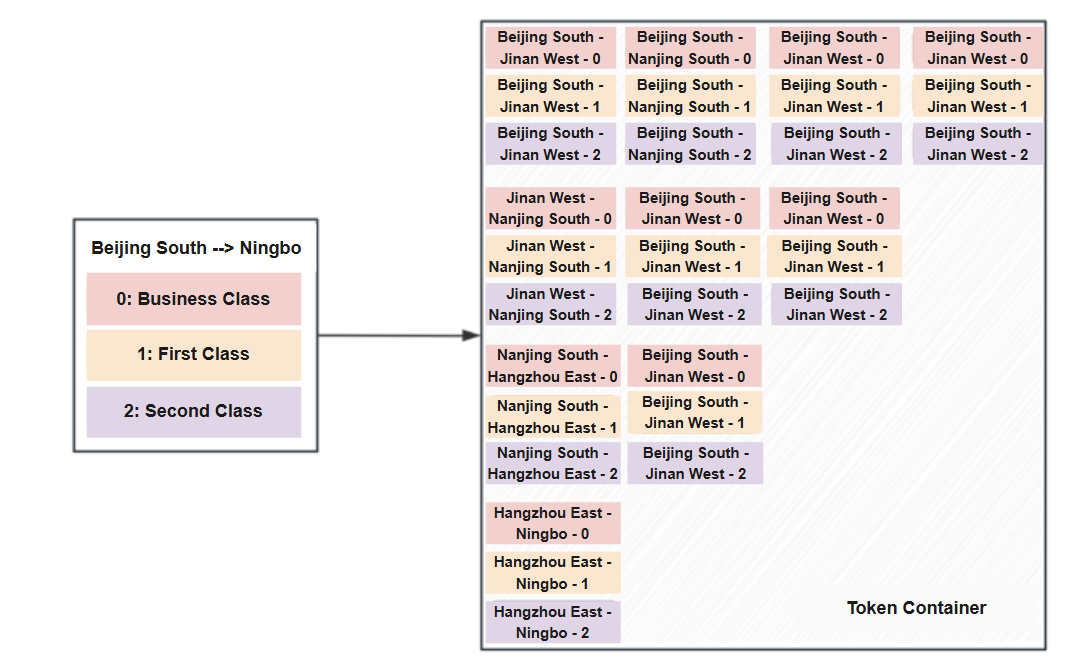}
        \captionof{figure}{Token Container Diagram}
        \label{fig:1}
        \end{center}

If a user wants to buy a ticket, the token container needs to be deducted first. Successful deduction means there are enough tickets left. Failure to deduct results in the request being rejected. This allows only those with the required number of tickets to receive tokens and enter the system, achieving rate limiting. Furthermore, because Redis uses a single-threaded model, the token deduction operation uses the INCRBY statement, which is atomic, thus ensuring that only the required number of tokens are sold, solving the problem of overselling inventory. Deducting tokens from the token container has no impact on the remaining ticket cache. Only after modifying the seating table will the remaining ticket cache be monitored and reduced via Canal to ensure data consistency between the database and the cache. Figure 4.5 illustrates the principle of the token rate limiting algorithm in solving the overselling problem and implementing rate limiting.

        \noindent
        \begin{center}
        \includegraphics[width=0.85\linewidth]{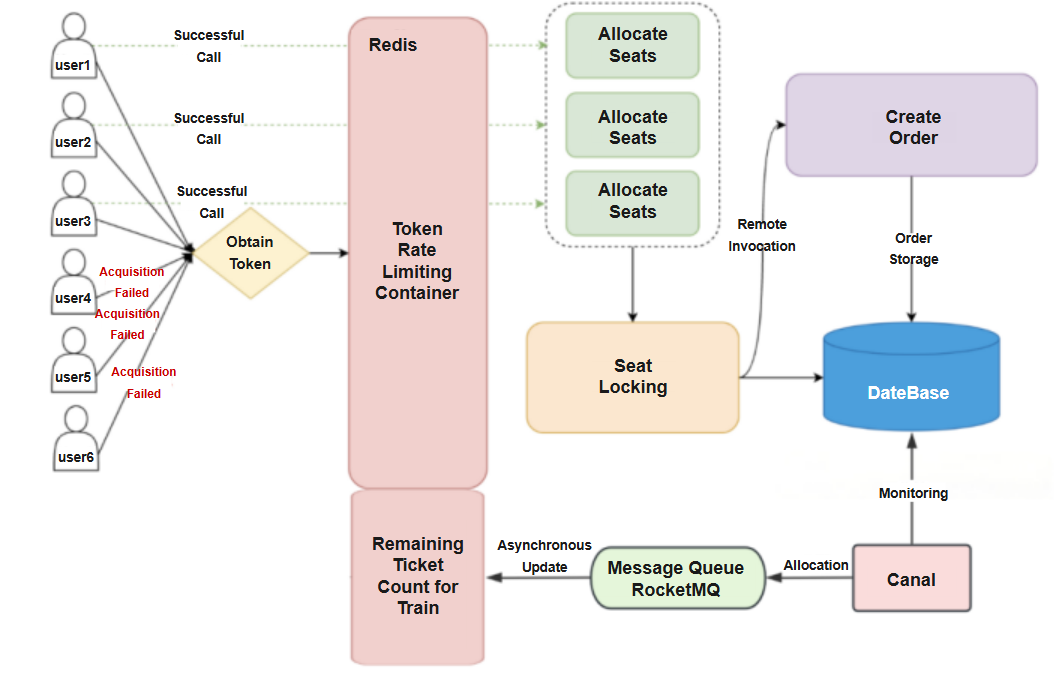}
        \captionof{figure}{Principle of Token-Based Rate Limiting}
        \label{fig:1}
        \end{center}

\subsection{4.2 Function Implementation}

\subsubsection{(1) Membership Services}
\vspace{12pt}

The user registration interface adopts a responsive form layout, including required fields such as username, password, ID card number, and mobile phone number. The front-end verifies the compliance of the 18-digit ID card format through a real-time validation mechanism and performs 11-digit numeric regular expression matching on the mobile phone number. The encryption module intercepts the SQL statements in the user request, parses them, and then encrypts the fields that need to be encrypted according to pre-set encryption rules. This system chooses the AES encryption algorithm.

After successful registration, users can log in by entering their username, phone number, or email address. Each user can also choose to add multiple passengers to purchase tickets for multiple people.

\subsubsection{(2) Ticketing Service}
\vspace{12pt}

After logging in, users can select multiple screening options to search for train routes based on their own needs. After selecting the train number, passenger, ticket type, seat, and other information, the ticket purchase process is complete. You can then submit the order and wait for the system to proceed with the next steps.

\subsubsection{(3) Order Service}
\vspace{12pt}

After a user submits an order, the system generates an order and temporarily locks the purchased seat. This system uses a customizable snowflake algorithm to generate a distributed ID and appends the last six digits of the user's ID to generate the order number \cite{39}. Clicking "Pay Order" will invoke the payment service. After completing the entire process, users can locate their orders within the Order Management module, under the section for Untraveled Orders.

\subsubsection{(4) Payment Services}
\vspace{12pt}

After the user clicks to pay for the order, they select a payment method. The system provides Alipay as the payment method. After successful payment, the system receives a successful payment notification from Alipay, the order status changes from unpaid to paid, and the seat changes from temporarily locked to successfully deducted.

\subsubsection{(5) View Tickets}
\vspace{12pt}

This can be achieved by binding the passenger's ID card number to the ticket order number and storing this binding mapping in a data table. After the passenger creates a personal account, they only need to enter their ID card number. The system will then route the data to the order number in the routing table, match the corresponding order details based on the order number, and view their ticket. The implementation idea and the corresponding page are shown in Figures 4.6, respectively. 

    \noindent
        \begin{center}
        \includegraphics[width=0.85\linewidth]{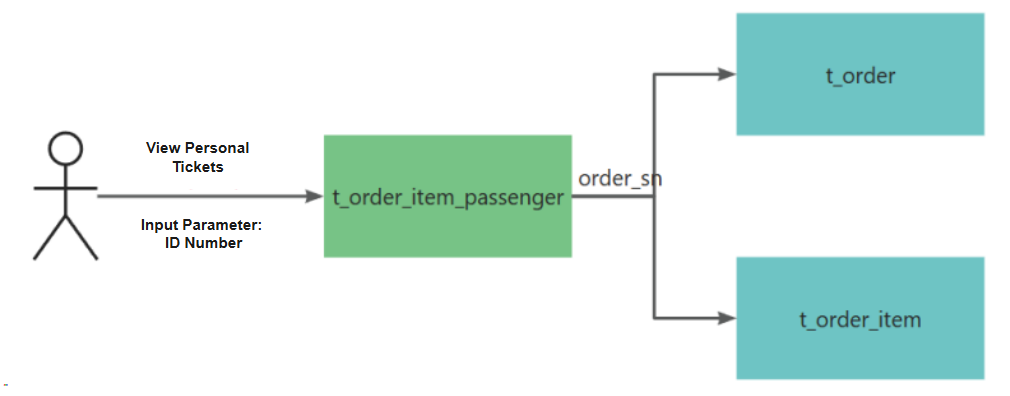}
        \captionof{figure}{Implementation Schematic}
        \label{fig:1}
        \end{center}
        
\subsection{4.3 Testing and Evaluation}
\vspace{0pt}

After the system's functionality and security design are implemented, it's necessary to scientifically test the performance metrics of the core functional interfaces frequently accessed by users, and then evaluate and analyze the generated test reports. The purpose of performance testing is to obtain the system's performance metrics to see if it meets the design requirements. Core metrics include response time, throughput, and request error rate. Typically, testing follows a specific strategy, applying pressure to the system within a reasonable range. After the testing process is complete, a data report is obtained. Analyzing this report can identify performance bottlenecks, find directions for performance optimization, and determine the system's suitability by simulating access demands from users of different scales.

This paper uses the Apache JMeter testing tool, specifically designed for testing software functionality and performance. Before testing, the system's environment configuration was modified, setting the JVM's maximum and minimum memory parameters to 4000MB, and the Tomcat server's thread pool parameters to a maximum of 300 threads, a minimum of 300 idle threads, a maximum concurrent connection count of 100,000, a maximum single-request throughput of 1000, and a maximum request queue length of 1000. This test was deployed on a local virtual machine. To reduce the pressure on the local server, a phased stress test scheme was adopted. The relevant parameters are as follows.

\begin{itemize}[noitemsep, leftmargin=*]
\item Input the final number of threads to be started in this test plan as 100.
\item After the test begins, threads will not start immediately. A 5-second pause is required before starting the first thread.
\item Set the initial number of threads to 10.
\item After the first thread starts, start 10 threads every 30 seconds, but these 10 threads must all start within 5 seconds.
\item After the number of threads reaches the maximum of 100 set in this test plan, continue running for 60 seconds. If system resources are exhausted, the test can be terminated early.
\item After running for 60 seconds, stop 5 threads per second until all started threads have stopped.
\end{itemize}

The test data generated after the performance test is completed is analyzed. The meaning of each column in the data table of the generated test report is as follows. Label is the name of each test that has been configured in the test environment. Sample is the total number of simulated requests sent to the system during the entire test process. Error rate is the percentage of errors in the total simulated requests sent to the system during the test. Throughput is the number of requests processed per second, which reflects the performance of the application. Each of the remaining columns is a statistical indicator of request response time, all in milliseconds. Finally, the receiving and sending speeds are the total amount of data transmitted per second. 

\begin{table}[htbp]
\centering
\caption{Performance Test Data of Train Query Interface}  
\label{tab:train_query}
\adjustbox{max width=\textwidth}{  % 自动调整表格宽度
\begin{tabular}{|l|c|c|c|c|c|c|c|c|c|c|c|}
\hline
\thead{Label} & 
\thead{Sample} & 
\thead{Average\\ Value} &   
\thead{Median} & 
\thead{90th\\ Percentile} & 
\thead{99th\\ Percentile} &   
\thead{Minimum\\ Value} & 
\thead{Maximum\\ Value} & 
\thead{Anomaly\\ Rate} & 
\thead{Through-\\ put} & 
\thead{Receive\\ KB/sec} & 
\thead{Send\\ KB/sec} \\
\hline
\thead{HTTP\\ Request} &   
5729 & 31 & 37 & 39 & 50 & 13 & 122 & 0.00\% & \makecell{817\\ /sec} & 2581.06 & 211.24 \\
\hline
TOTAL & 
5729 & 31 & 37 & 39 & 50 & 13 & 122 & 0.00\% & \makecell{817\\ /sec} & 2581.06 & 211.24 \\
\hline
\end{tabular}
}
\end{table}

\begin{table}[htbp]
\centering
\caption{Performance Test Data of Ticket Purchase Interface}
\label{tab:train_query}
\footnotesize % 使用更小的字体
\adjustbox{max width=\textwidth}{ % 自动缩放表格到页面宽度
\begin{tabular}{|l|c|c|c|c|c|c|c|c|c|c|c|}
\hline
\thead{Label} & 
\thead{Sample} & 
\thead{Average\\ Value} & 
\thead{Median} & 
\thead{90th\\ Percentile} & 
\thead{99th\\ Percentile} &
\thead{Minimum\\ Value} & 
\thead{Maximum\\ Value} & 
\thead{Anomaly\\ Rate} &
\thead{Through-\\ put} & 
\thead{Receive\\ KB/sec} & 
\thead{Send\\ KB/sec} \\ 
\hline
\thead{HTTP\\ Request} & 
16672 & 65 & 59 & 65 & 231 & 21 & 231 & 2.01\% & \makecell{265\\ /sec} & 76.78 & 367.30 \\ 
\hline
TOTAL & 
16672 & 65 & 59 & 65 & 231 & 21 & 231 & 2.01\% & \makecell{265\\ /sec} & 76.78 & 367.30 \\
\hline
\end{tabular}
}
\end{table}

The performance test data for the train query interface is shown in Table 4.1, and the performance test data for the ticket purchase interface is shown in Table 4.2. The test results show that the system performs well in terms of stability and basic performance under high concurrency. The average response time of the train query interface is 31 milliseconds, and the throughput is 817 requests per second, with no request anomalies. However, 99\% of requests still take 50 milliseconds, with extreme cases reaching 122 milliseconds, indicating a potential throughput bottleneck. The average response time of the ticket purchase interface is 65 milliseconds, and the throughput is 265 requests per second. Due to local resource limitations, this interface experienced a request anomaly rate of 2.01\% due to thread pool resource exhaustion. In the future, this system will be deployed on multiple servers, and the service will be horizontally scaled by increasing the number of instances to improve system throughput and availability. Simultaneously, measures such as tracking and analyzing time-consuming operations and optimizing inefficient SQL statements will be implemented to improve system performance. 

\section{5 CONCILUSION} 
\vspace{0pt} 

This paper presents the design and implementation of a stable, high-performance railway ticketing system based on a microservice architecture, specifically designed for high-concurrency scenarios. The system is built on Spring Cloud Alibaba and addresses several challenges through integrated solutions. At the data layer, ShardingSphere implements horizontal sharding of user and order tables, while Redis caches frequently accessed data, reducing the response time of core interfaces from 200 milliseconds to less than 40 milliseconds. To ensure consistency between the cache and the database, the system uses Canal to monitor Binlog log changes, and RocketMQ handles asynchronous updates. Furthermore, the system implements an atomic token bucket algorithm based on Redis for ticket inventory deduction, complemented by Sentinel for traffic control, effectively preventing overselling of tickets. The final tests are conducted after deploying the service on a local virtual machine. The results shows that the system's train query interface achieved a throughput of 817 queries per second with an average response time of 31 milliseconds, the ticket purchase interface had an average response time of 65 milliseconds and a throughput of 265 queries per second, with zero overselling of tickets, verifying the reliability and stability of the system design. Furthermore, an innovative approach was proposed, using a customizable length Snowflake algorithm to replace UUIDs for generating distributed globally unique IDs. This results in monotonically increasing IDs. Compared to traditional UUID schemes, the new generation algorithm significantly improves database index write performance while requiring less storage space for the generated IDs, effectively supporting order storage and retrieval in high-concurrency environments.

\section*{ACKNOWLEDGEMENTS}
\vspace{0pt}

During the preparation of this paper, we used ChatGPT for language refinement and translation assistance.

\end{document}